\documentclass[10pt,a4paper]{article}

\usepackage{graphicx}
\usepackage{fullpage}
\usepackage[hidelinks]{hyperref}
\usepackage{wrapfig}
\usepackage{framed}
\newcommand{\junk}[1]{}

\title{Coinami: A Cryptocurrency with DNA Sequence Alignment as Proof-of-work}
\author{
  \centering
  Atalay M. Ileri$^{1,*,\dag}$  \and Halil I. Ozercan$^{1,*}$ \and Alper Gundogdu$^{1}$ \and Ahmet K. Senol$^{1}$ 
  \and M. Yusuf Ozkaya$^{1,\sharp}$ \and Can Alkan$^{1,\ddag}$ \\
  \footnotesize
    ${}^1$ Department of Computer Engineering, Bilkent University, Ankara, Turkey\\
  \footnotesize
    ${}^\dag$ Current address: Department of EECS, Massachusetts Institute of Technology, Cambridge, MA, United States \\
  \footnotesize
    ${}^\sharp$ Current address: College of Computing, Georgia Tech, Atlanta, GA, United States \\
  \footnotesize
    ${}^*$ These authors contributed equally.
  \footnotesize
    ${}^\ddag$ Corresponding author: \url{calkan@cs.bilkent.edu.tr}.
}

\begin{document}
\date{}
    \maketitle
\begin{abstract}
Rate of growth of the amount of data generated using the high throughput sequencing (HTS) platforms now exceeds the growth stipulated by Moore's Law. 
The HTS data is expected to surpass those of other ``big data'' domains such as astronomy, before the year 2025. In addition to sequencing 
genomes for research purposes, genome and exome sequencing in clinical settings will be a routine part of health care. The analysis of such
large amounts of data, however, is not without computational challenges. This burden is even more increased due to the periodic updates to reference genomes, which
typically require re-analysis of existing data. 

Here we propose Coin-Application Mediator Interface (Coinami\footnote{All students involved 
in this project were undergraduates when they contributed. No graduate students were harmed during the making of
 this project. Yet.}) to distribute the workload for mapping reads to reference genomes
using a  volunteer grid computer approach similar to Berkeley Open Infrastructure for Network Computing (BOINC).
However, since HTS read mapping requires substantial computational resources and fast analysis turnout is desired, Coinami uses the HTS read mapping
as proof-of-work to generate valid blocks to main its own  cryptocurrency system, which may help motivate volunteers to dedicate more resources.
The Coinami protocol includes mechanisms to ensure that jobs performed by volunteers are correct, and provides genomic data privacy.
The prototype implementation of Coinami is available at \url{http://coinami.github.io/}.
\end{abstract}

\section{Introduction}

High throughput sequencing (HTS) technologies evolved very quickly since 2007~\cite{Metzker2010}, and now they are among the most powerful tools available for biological research.
We are now able to read the entire genome of a human individual in a few days for a fraction of  costs incurred by previous technologies~\cite{Mardis2008,Metzker2010}.
However, the volume of data generated by these platforms are enormous, leading to a picture where computational analyses represent the major bottleneck~\cite{Flicek2009,Treangen2012,Sboner2011}. 
For example, the Illumina HiSeqX platform can sequence the genomes of approximately 18,000 humans  a year,
at an estimated cost of \$1,500 per genome ({\url{http://www.illumina.com/systems/hiseq-x-sequencing-system.ilmn}). 
This corresponds to about 2 petabytes of data per year, per sequencing center. Considering the fact that there are many genome centers that either already have, or will purchase this system, the amount of data generated each year will increase to hundreds of petabytes to exabytes.

The computational analyses of such data involves multiple steps, but the main bottleneck is to find the potential locations of short stretches of DNA sequences in a reference genome. 
This step, called {\it read mapping}, usually takes ~30 CPU days per human genome~\cite{Li2009a,Li2013,Alkan2009,Langmead2009,Hach2010,Xin2013} (see~\cite{Fonseca2012} for a review of aligners).
The computational burden of read mapping is monotonically increasing not only because of the growth of  data to be analyzed, but also
because of updates in the reference genome assemblies. For instance, the human reference genome is updated every 3 to 4 years that fixes assembly mistakes, and adds either 
new sequences that are found in the genomes of
newly sequenced individuals, or ``alternative haplotypes'' that are frequent variations from the existing reference sequence. Such drastic changes in reference genomes usually necessitate
 remapping of the existing data to enable more accurate characterization of genomic variants.
Thanks to the {\it embarrassingly parallel} nature of this problem clusters are typically used.
However, building clusters that are large enough to handle hundreds of petabytes of data is not feasible. 
Therefore,  volunteer grid computing technologies, on the other hand, offer a promising alternative to large clusters and data centers. 
Although there are certain distinctive characteristics of mapping reads generated by different platforms, given the popularity of Illumina as today's  most popular sequencing platform, 
 the remainder of the paper will assume that the genomes are sequenced using the Illumina technology.

Volunteer grid computing was made popular by the Berkeley Open Infrastructure for Network Computing (BOINC) platform, and specifically its Search for Extraterrestrial Intelligence at Home (SETI@home) 
project (\url{http://setiathome.ssl.berkeley.edu/}). BOINC volunteers choose a scientific problem to work on, download data from the server, solve the problem, and upload results back to the server. To reduce the burden on the volunteers, the BOINC clients can be set to run only when the computer is idle (i.e. ``screen saver mode''). There are a number of bioinformatics applications ported to the BOINC platform such as Rosetta@home for protein structure prediction (\url{https://boinc.bakerlab.org/}), 
and FiND@home (\url{http://findah.ucd.ie/}) for docking simulations on malaria proteins.
Although it is also possible to write HTS read mapping applications for BOINC, such applications must provide data security to prevent both ``leak before publication'' and genomic privacy, and protection against malicious users (i.e. volunteers that deliberately upload incorrect results). Furthermore, HTS read mapping ``assignments'' (i.e. the sequence data) continuously grow as the volume of 
 data required to be processed grows.
It is obvious that timely analysis requires  substantial computational resources (CPU, RAM, disk space, and network bandwidth), which may make it difficult to motivate volunteers. 
One path to increasing volunteer motivation to dedicate resources runs, arguably,  through incorporating cryptocurrencies within the distributed grid computing for HTS read mapping.

The cryptocurrencies were first introduced by Wei Dai in 1998\cite{Dai1998}, and 
 received considerable attention from the public after the first decentralized cryptocurrency (Bitcoin) was released in 2009 by an unknown person with 
    pseudonym Satoshi Nakamoto\cite{Nakamoto2008}. 
The motivation behind Bitcoin  was to create a distributed currency, which is not dependent on any controlling authority. 
   On the footsteps of Bitcoin's success, a plethora of cryptocurrencies were proposed including Litecoin~\cite{litecoin}, Peercoin~\cite{peercoin} 
and Dogecoin~\cite{dogecoin}. More recently, another proof-of-work scheme, named Cuckoo, is introduced that finds small cycles in large random graphs~\cite{cuckoo}.
These cryptocurrencies are also maintained by the ``miners'', but the power of their mining network is negligible compared to that of Bitcoin's. 

Although there are some differences between cryptocurrencies, all of them are composed of two interconnected processes called Mining and Transaction. In a nutshell, mining
refers to the creation of new ``digital money'' termed {\it coin}, that are recorded by a public ledger called {\it blockchain}. Mining involves a computationally difficult process called ``proof-of-work'', while transaction refers to an exchange of blockchains between users.  In Section~\ref{sec:protocol} we provide some insight into cryptocurrency protocols in detail in the context of Bitcoin - the most widely used cryptocurrency.

As of January 2016, Bitcoin mining network's total computation power has reached approximately 710 PetaHash/s\footnote{\url{http://blockchain.info/stats}}.
    In comparison, world's most powerful scientific computation grid -BOINC- 
    boasts a computation power of 11.22 PetaFLOP/s\footnote{\url{http://boinc.berkeley.edu/}}.  
 Although two units  are not directly comparable as hashing uses integer operations while BOINC's power is measured in floating point operations, 
even the arbitrary yet not too radical scaling of 1 to 10 (assuming floating point operations are 10X ``harder'' than integer operations) posits Bitcoin network almost as powerful as the BOINC network.

The computation power of the Bitcoin network is mainly used to maintain the currency’s integrity by ensuring that the blockchain creation is always a difficult task. The difficulty stems from the proof-of-work, which entails finding a number called the {\it nonce}. Nonce computation becomes harder and harder as the {\it difficulty target} is increased after the creation of every 2016 blocks. This effectively limits the amount of blockchains generated by the miners, preventing devaluation of the money as more blockchains are mined. The proof-of-work schemes within Bitcoin and other cryptocurrencies are effective solutions for the security of the cryptocurrency systems, yet serve no other practical purpose.

Here we propose Coin-Application Mediator Interface (Coinami) where such computation power will be used for scientific computation as well as integrity purposes. We chose to use DNA sequence alignment problem as the proof-of-work in our initial implementation. Briefly, Coinami is a three-level multi-centric system that distributes HTS alignment problems to volunteers (or, {\it miners}). The miners download problem sets from the middle level autorities, that are in turn certified by the root authority, map the HTS reads to a reference genome, and send the results back to the middle level authority for verification (Section~\ref{sec:coinami}).
However, the proof-of-work is decoupled from the rest of the system, making Coinami easily adaptable to other scientific problems that require substantial computational resources.

\section{Current cryptocurrency systems}
\label{sec:protocol}
In this section we provide a brief introduction to cryptocurrencies based on the Bitcoin protocol. 

\subsection{Transactions}
{\it Transactions} 
 represent money exchange between two parties. Simply, they are records of participants, input amount and output amount. Unlike physical money (i.e. traditional currencies such as Dollar, Euro, etc.), cryptocurrencies can easily be copied before spending them. Therefore, all transactions should be recorded to prevent double spending. Every time a transaction occurs, the original (input) coins are destroyed and new (output) coins are generated. This scheme ensures that every coin can be spent only once. Each transaction contains receiver's public key and is signed by sender's private key. This adds the information of the spender and the receiver to the transaction itself. Once a transaction created, it is broadcast to the network for inclusion in a block. Inclusion in a block makes a transaction valid.

There is a special transaction called {\it coinbase transaction}, which does not have any input coins. It is the first transaction of every block and it contains the block generation reward for the miner.

\subsection{Mining}
    Generation of a new block is called {\it mining}. The main purpose of mining is validating and recording the transactions. A fresh-minted block contains the hash of the previous block in chain, a set of transactions and a {\it nonce}, which is a 32 bit integer used for altering the block’s hash value 
 (Figure~\ref{fig:bitcoin}A).
A block is considered to be {\it valid} if its hash contains a predetermined number of zeros at the beginning, which denotes the {\it difficulty} of the generated block. 
Upon the generation and publication of the block, miners are awarded with some new coins. These new coins are included in the block as the first transaction and this transaction is called
{\it generation transaction} or {\it coinbase transaction}.

 	\begin{figure}[ht!]
 	  \begin{minipage}[t]{0.45\textwidth}
          \begin{center}
 	      A)\\ 
              \includegraphics[width=0.8\textwidth]{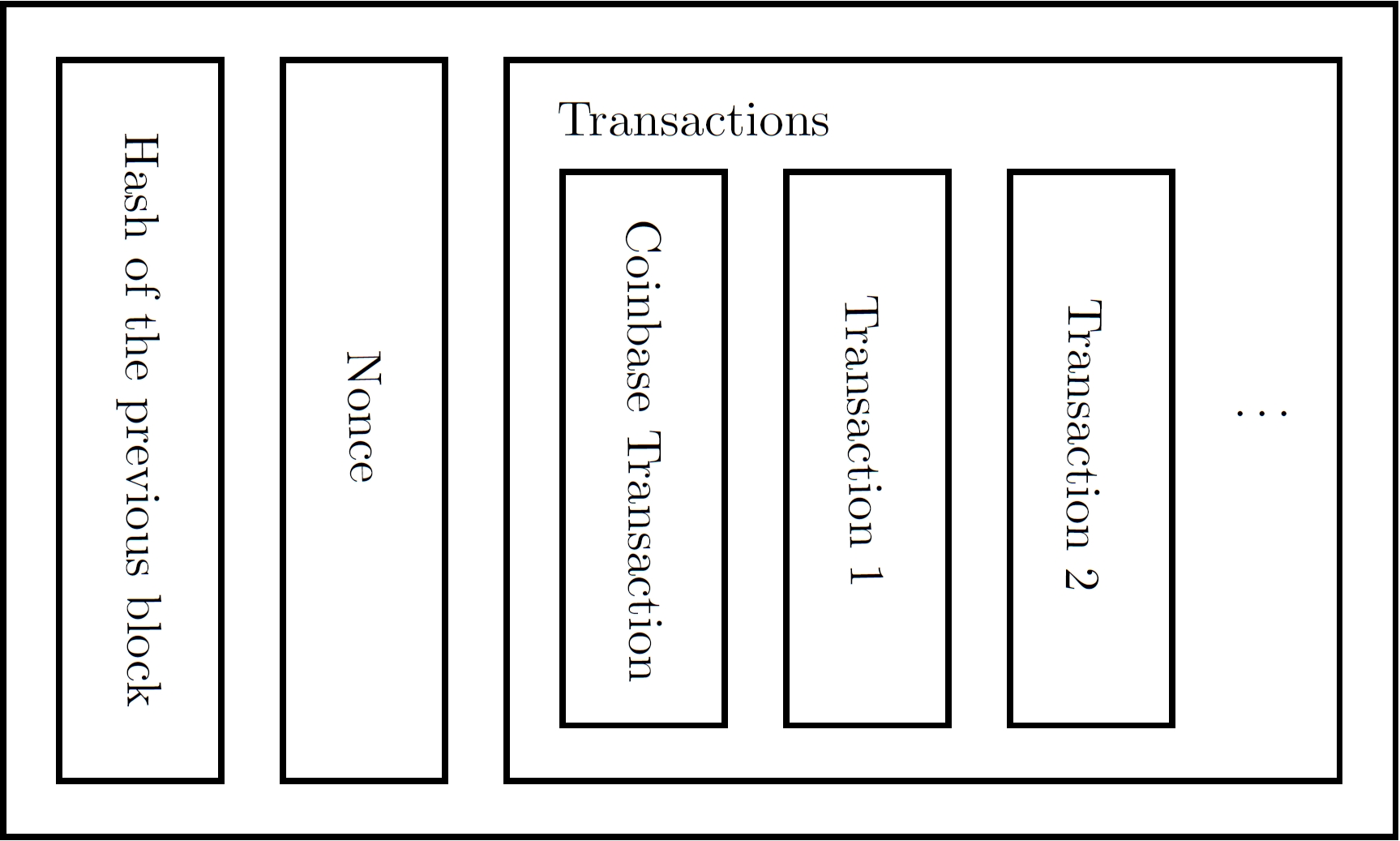}
              \vspace*{0.7cm} 
            \end{center}
          \end{minipage}
          \hfill
 	  \begin{minipage}[t]{0.45\textwidth}
            \begin{center}
 	      B)\\ 
              \includegraphics[width=0.8\textwidth]{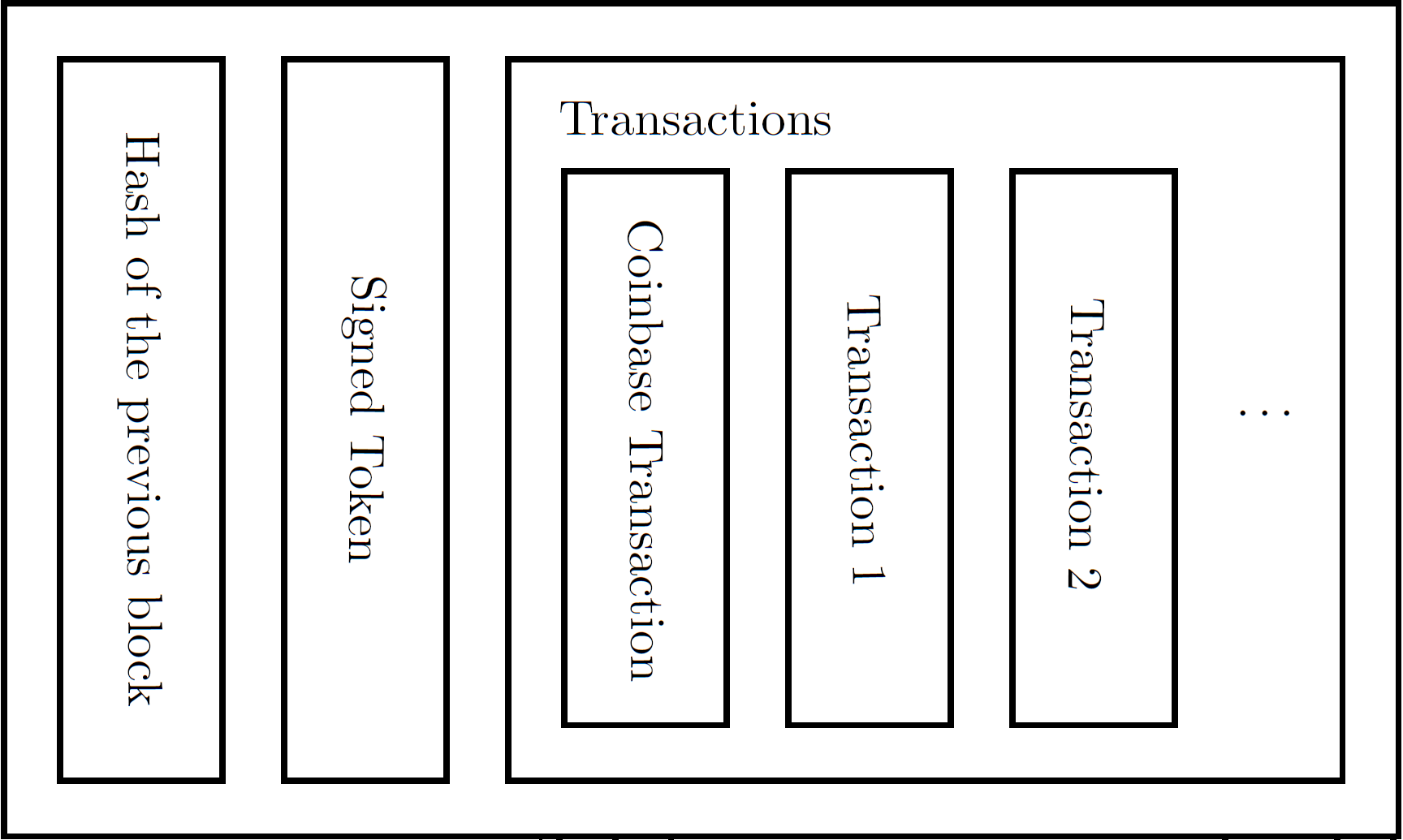}
            \end{center}
          \end{minipage}
 	  \caption{{\small A) Contents of a Bitcoin block. 
            B) Coinami block. We modified the Bitcoin blockchain by replacing the {\it nonce} with a {\it signed token}, which is generated by the authority. The remainder of the block is the same
            with that of Bitcoin, as it includes the hash of the previous block and the transaction history.}}
          \label{fig:bitcoin}
          \line(1,0){454}\\
 	\end{figure}


Computational power of a mining network can change dynamically. The difficulty index is introduced to prevent fluctuations in coin throughput due to these changes, 
and is updated after the generation of every 2016$^{th}$ block. The Bitcoin mining protocol assumes that it takes approximately one week to generate 2016 blocks, and adjusts the difficulty level based on the actual time.

In practice, mining is performed by taking a potential block’s hash using the SHA-256 hash function to find a value with predetermined difficulty. 
Alteration of the hash value is achieved by changing the value of nonce. This is done by trying different nonce values in a brute-force fashion; the result is called 
{\it proof-of-work}.

\section{Proposed system}
\label{sec:coinami}
Since virtually all computational power required in the process is used for the calculation of proof-of-work, we integrate our solution in that portion of the protocol. The two important properties of the proof-of-work scheme are 1) maintaining the difficulty of generation, and 2) validity of blocks. Therefore, the new proof-of-work scheme should keep these properties in place. 
Most current cryptocurrencies, including Bitcoin, are completely decentralized, as the assignments for the proof-of-work can be generated independently by the miners as long as they meet the system's difficulty level. However, for the proof-of-work in Coinami to be {\it useful}, the assignments for the proof-of-work need to 
have real and practical value. Therefore unlike major cryptocurrencies, Coinami is not completely decentralized due to the availability and generation of HTS data. 
Instead, Coinami
has a three-level multi-center structure, where one {\it root authority} tracks and validates middle level {\it authority servers}
that supply HTS data to the system and checks for validity of alignments, and the third level is composed of {\it miners} (Figure~\ref{fig:structure}). 
The root authority should be trusted by the entire system to validate only ``trustable'' authority servers, which can be major sequencing centers. In the remainder of the paper, we 
refer to middle level authority servers as {\it authorities}.

\begin{figure}[ht!]
  \centering
  \includegraphics[width=0.7\textwidth]{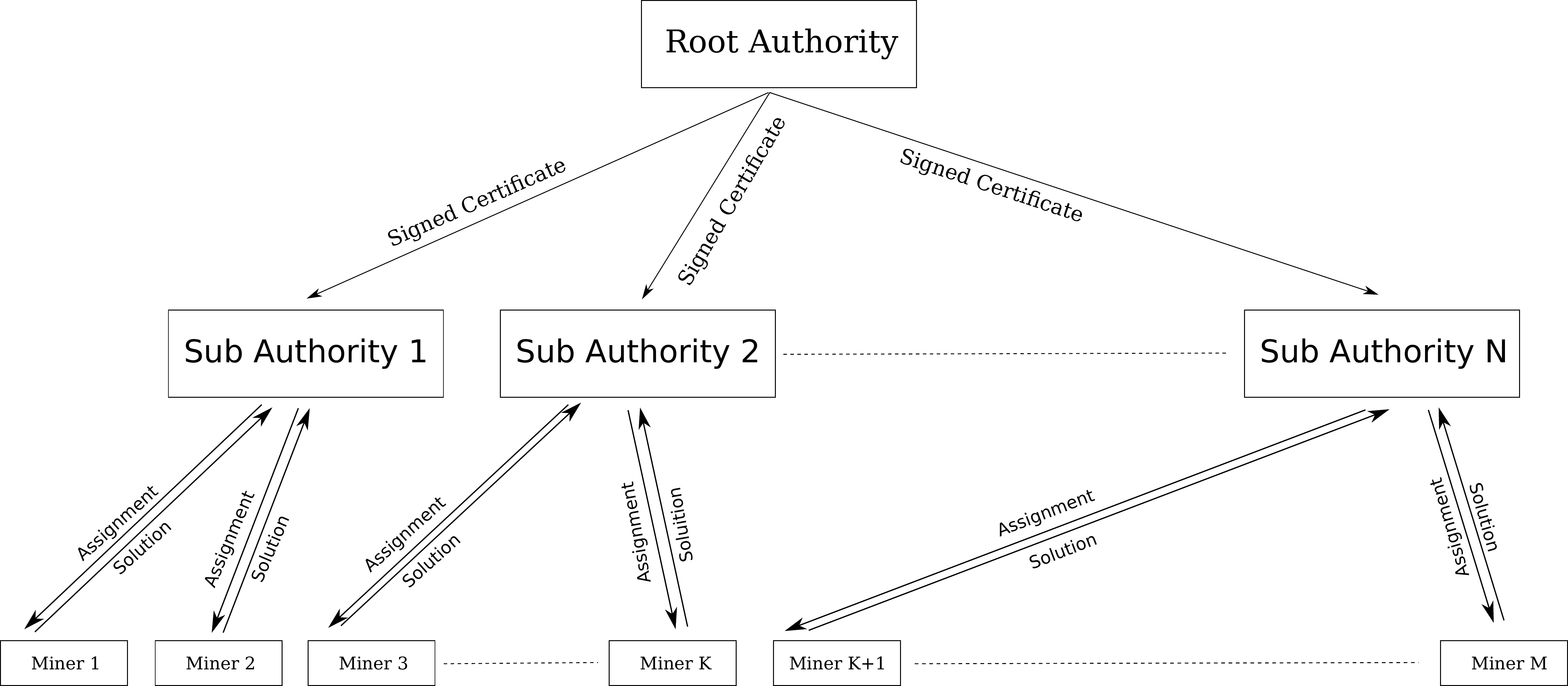}
  \caption{Three level structure for the Coinami network. The single root authority issues certificates to the sub authorities. The sub authorities (middle level) sends assignments to multiple miners, and validate the results they receive. If the alignments are valid, they sign the blockchains and return to the miners. A miner can work on assignments from multiple sub authorities.}
  \label{fig:structure}
  \line(1,0){454}\\
\end{figure}

    \begin{figure}[!h]
    	\centering
    	\includegraphics[width=0.88\textwidth]{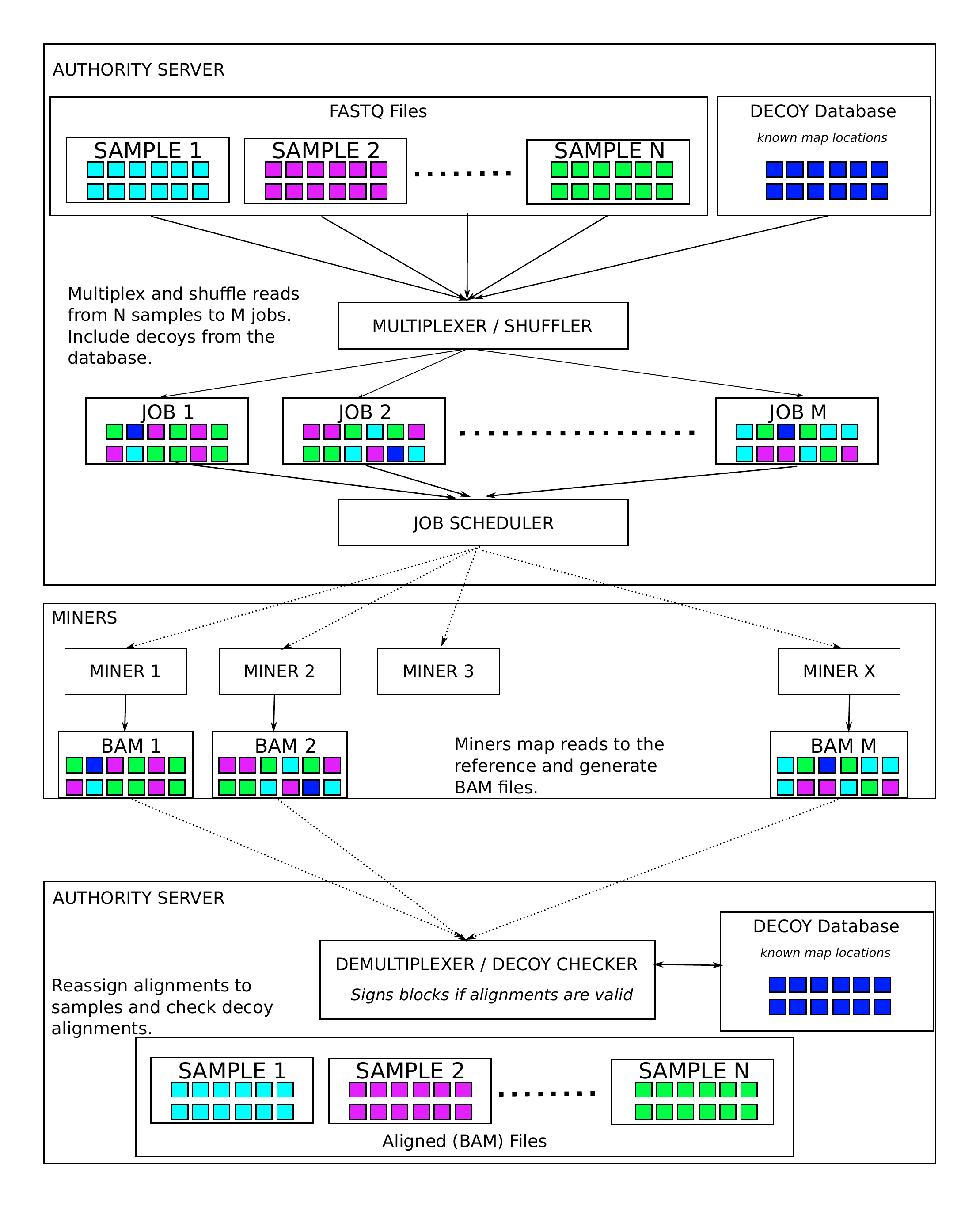}
    	\caption{{\small Coinami workflow. The authority hosts genomes of $N$ samples in the form of FASTQ files, and a database of decoy reads with known mapping locations. The decoy database needs to be generated once for each reference genome and read mapper combination. To create the assignments, the authority multiplexes and shuffles reads from multiple samples and includes decoys that constitute
 5\% of the assignment. The miners requests jobs from the authority to to work on, which are tracked by a job scheduler that also imposes ``deadlines'' on the assignments to prevent 
        deadlocks. The miners download the assignments, map the reads to the reference genome specified by the exchange protocol between the authority and the miners, and send the BAM files back to the
        authority. Finally, the authority reassigns the alignments to $N$ samples and simultaneously checks decoys to verify the results. If the decoys are aligned to their predetermined location, the BAM file is considered to be valid, and the authority signs the block and returns it back to the corresponding miner.}}
        \label{fig:coinami-workflow}
          \line(1,0){454}\\
    \end{figure}

Figure~\ref{fig:coinami-workflow} summarizes the Coinami workflow.
We adapted our scheme from Adam Back's Hashcash~\cite{hashcash} protocol. Below, we provide the Coinami protocol in the context of the middle level authorities and the miners. 

\subsection{Authorities}

The root authority assigns certificates to the middle level authorities, {\it validating} the authorities and allowing them operate within the Coinami network. 
The certificates and their corresponding private keys will be used in {\it token} signing.

Authorities fulfill two main roles in our system: 1) they {\it inject} new assignments (i.e. ``alignment problems'') 
into the system, and 2) check for the validity of the results to prevent counterfeit. If the results are valid, the authority signs the block by adding a signature token and returns it to the miner. For the validity test, the authority injects {\it decoy reads} into the assignment. The decoy reads will constitute 5\% of each assignment, and will allow quick comparison of the accuracy of alignment results, as theirs are pre-calculated. Note that in a set of 1,000 reads, there will be 50 decoys, and the probability of guessing the decoys correctly is $\frac{1}{{1,000\choose 50}} \leq 1.06\times 10^{-85}$. To prevent errors in downstream analyses of the alignment results for genomic variation discovery, the authority discards alignments for the decoy reads after verification of the results. Additionally, the decoy reads need to be indistinguishable from the rest of the assignment. In the case a read aligner that selects a random location for multi-mapping reads~\cite{Li2009a,Li2013} is used, 
either the decoys need to be selected among uniquely mappable reads (which may pose an attack threat), or all possible locations should be considered as correct (i.e. the {\tt XA} field as reported by BWA-MEM).
Alternatively, a deterministic mapper such as Bowtie2~\cite{Langmead2009} or mrFAST~\cite{Alkan2009,Xin2013} can be used.
As an extra step to enhance data privacy, the authority mixes reads from multiple genomes together with the decoys into a single assignment (i.e. FASTQ file).
Multiplexing data from the genomes of different individuals makes reconstruction of an individual's sequence impossible even if miners cooperate.
Note that if the assignment contains $2n$ reads from two data sets, then there are $2^n$ different ways of grouping them into two size $n$ subsets. 
The size of the assignments (i.e. number of reads) determines the difficulty of mining.
The authority then encrypts the read names using AES encryption~\cite{Daemen1998} to prevent the miners from deducing the decoys and individual genomes (Figure~\ref{fig:assignment}). 
Note that the authority is the only entity that can decrypt read names using its own key.
Then, a job scheduler
adds ``deadlines'' to the assignments to prevent deadlocks, and makes them available for downloading by the miners (Figure~\ref{fig:coinami-workflow}).
The assignment also includes metadata for the reference genome and the read mapper to be used along with its mapping parameters. We note that the pre-calculated alignments for the decoys
will need to be updated for different aligners and parameters, however, in practice only a few read mappers and their default parameters are widely used.
After the miner sends the results (i.e. BAM file~\cite{Li2009b}), the authority separates
 the alignments from multiple genomes and verifies (and discards)  decoys simultaneously by simply scanning the result file. 
Once the authority is satisfied, it signs and sends the block back to the miner.

 	\begin{figure}[!ht]
            \begin{center}
                  \includegraphics[width=\textwidth]{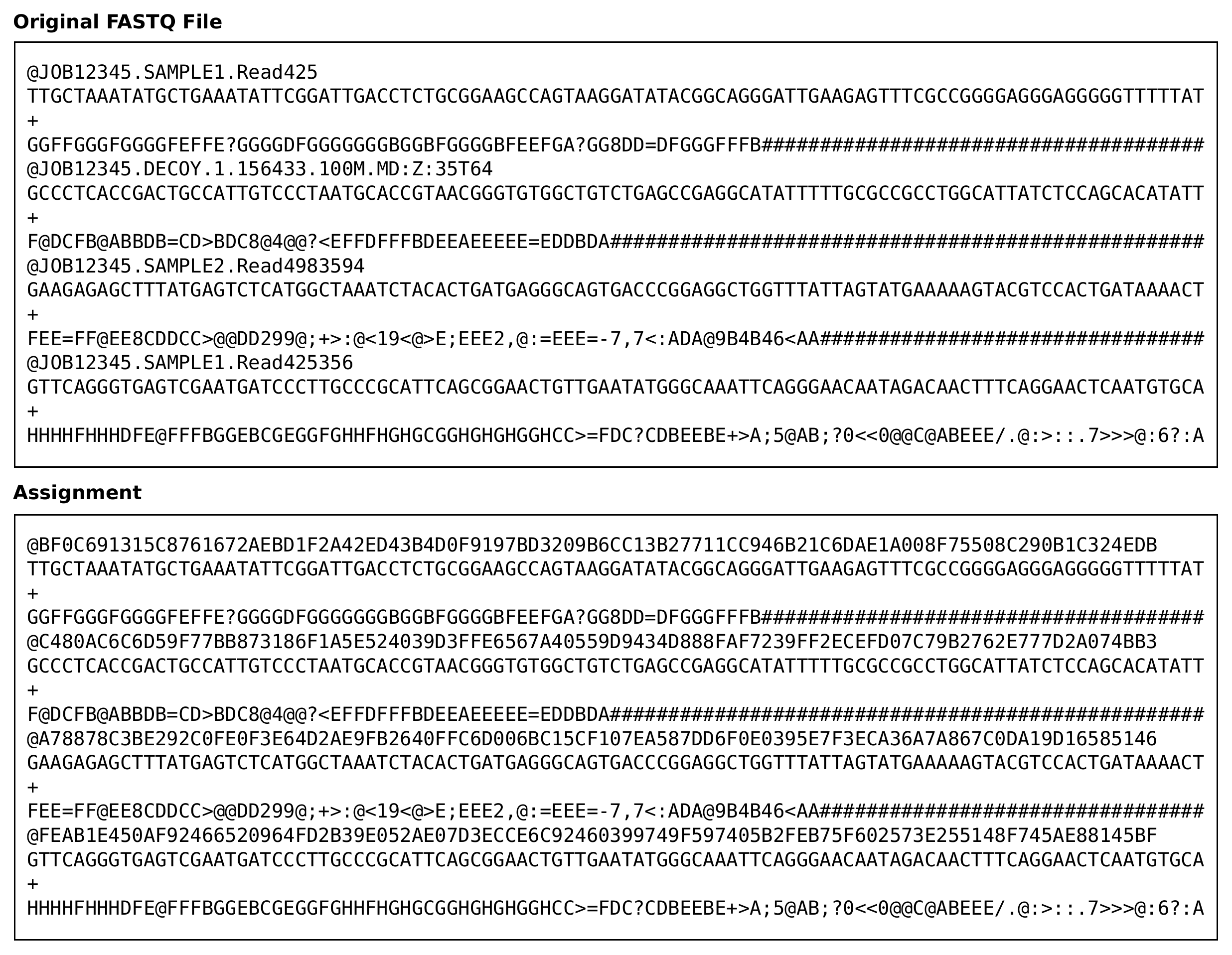}
            \end{center}
 		\caption{ {\small Sample assignment. For simplicity we only present one of the paired-end reads.
                  The authority first generates a FASTQ file by mixing and shuffling reads from multiple samples and inserting decoy reads. 
                  For simplifying demultiplexing of the results, reads are renamed and labeled with sample IDs. The decoy read names also include the mapping information that can 
                  be directly compared with the miner-reported BAM file to avoid database search. 
                  In this example, the decoy read is pre-aligned to coordinate 156,433 in chromosome 1, with CIGAR and MD fields~\cite{Li2009b} 100M and MD:Z:35T64, respectively.
                  We also add a job ID as a prefix to read names (JOB12345 in this example), which acts as a salt for the encrypted read name. This way, even if the same
                  decoy is used in different job/assignment, it receives a different encrypted read name, preventing a potential attack.
                  However, this information needs to be hidden from the miners to prevent decoy and sample identification. Therefore, the authority encrypts the read names with
                  its private key, and applies base64 encoding to represent encrypted read names in ASCII format. Read name encryption also generates the same read name for the second end of
                  the paired-end reads.
                }}
                \label{fig:assignment}
          \line(1,0){454}\\
 	\end{figure}

\subsection{Miners}
A miner requests jobs from an authority and the authority sends the miner the next available assignment. 
The miner then maps the reads within the assignment to the reference genome using the designated mapper and its parameters, and generates a BAM file. 
The BAM file is sorted and indexed for easier processing at the authority side. However, the miner does not remove duplicates as done in most
analyses, simply because the assignment contains reads from multiple genomes along with decoys. Therefore duplicate removal process will likely discard reads 
that are common across different genomes and decoys by chance, thus invalidating the alignment. 

The miner sends the sorted and indexed BAM file back to the authority for validation. As described above, if the decoys are aligned to their correct positions, the authority signs and sends
the block (Figure~\ref{fig:bitcoin}B) back to the miner. 
Upon receiving the validated block, the miner spreads it within the network. Other miners validate the new block by checking the signature of the authority, 
and other basic cryptocurrency block properties such as the hash of the previous block hash and the transaction history..

\subsection{Implementation} 
We implemented a prototype for Coinami authority servers in C++, and the interface for miners including a graphical user interface for wallet in Python. The current version runs on  Linux operating system. 
It uses BWA-MEM~\cite{Li2013} for read mapping, SAMtools~\cite{Li2009b} for SAM to BAM conversion and BAM sorting, and the {\tt libstatgen}\footnote{\url{http://genome.sph.umich.edu/wiki/C\%2B\%2B_Library:_libStatGen}} library for BAM processing (i.e. decoy identification and removal, and sample recovery).
We used the Crypto++ library\footnote{\url{https://www.cryptopp.com/}} to implement hashing, signing, and encrypting/decrypting read names.
We modified the BasicCoin\footnote{\url{https://github.com/zack-bitcoin/basiccoin}} project to implement the Coinami blockchain.
Coinami components for authorities and miners are available at {\url http://coinami.github.io/}. 

\subsection{Processing power estimates}

In this section we provide an estimate of the processing power of the Coinami network. First, we ran two sample assignments with 100 bp reads using three different CPUs (Table~\ref{tab:gflops}), and compared the run
times with the estimated processing power in billion floating point operations (gigaflop; GFLOP) per second (GFLOP/s). We used the CPU performance table provided by BOINC\footnote{\url{https://setiathome.berkeley.edu/cpu_list.php}} per core, and multiplied the run time in seconds with the estimated GFLOP/s values to estimate the number of floating point operations.  We then calculated the number of reads processed per gigaflop.  We note that, most of the assignment work is in fact integer operations, however, this analysis gives us a rough estimate for proof-of-work run times across different CPUs. Even if we consider the lowest performance, on the average, 628 reads are processed per GFLOP. Therefore, if Coinami network processing power reaches 10\% of that of BOINC's (i.e. 1 PetaFLOP/s), $\sim$658 million reads can be processed per second (if we omit the data transfer time), which corresponds to one genome sequenced at 22X coverage. Note that this estimate does not include the data transfer and authority processing times.

\begin{table}[ht]
\caption{Benchmarking results for number of reads processed per gigaflop.}
\begin{center}
\begin{tabular}{lrrrrr}
\hline
  \textbf{CPU}                              & \textbf{Reads} & \textbf{Time (s)} & \textbf{GFLOP/s} & \textbf{GFLOP} & \textbf{Read/GFLOP}  \\
\hline
  Intel$^{R}$ Core$^{TM}$ i7-3770 3.40GHz            & 1,000,000        & 195                & 4.4              & 858            & 1,165.50           \\
  Intel$^{R}$ Core$^{TM}$ i7-3770 3.40GHz             & 10,000,000       & 1,945               & 4.4              & 8,558           & 1,168.50        \\
\hline
  Intel$^{R}$ Xeon$^{R}$ E5-2643  3.30GHz            & 1,000,000        & 304                & 3.99             & 1,212.96        & 824.43          \\
  Intel$^{R}$ Xeon$^{R}$ E5-2643  3.30GHz            & 10,000,000       & 2,992               & 3.99             & 11,938.08       & 837.66         \\
\hline
  Intel$^{R}$ Xeon$^{R}$ E7- 4830 2.13GHz             & 1,000,000        & 698                & 2.28             & 1,591.44        & 628.36         \\
  Intel$^{R}$ Xeon$^{R}$ E7- 4830 2.13GHz             & 10,000,000       & 5,645               & 2.28             & 12,870.6        & 776.96       \\
\hline
  \end{tabular}
\end{center}
\small{Analysis of computation power vs. proof-of-work. We sampled two sets of 100 bp reads from the 1000 Genomes Project~\cite{1000GP2015} to generate assignments. As the proof-of-work,
we aligned the reads to the reference human genome using BWA-MEM~\cite{Li2013}, then converted SAM file to BAM and sorted the BAM file using SAMtools~\cite{Li2009b}. The run times reported in this 
table corresponds to the total time required to run both BWA-MEM and SAMtools using a single core.}
\label{tab:gflops}
\line(1,0){454}\\
\end{table}

\section{Future Work}
There are many yet unsolved problems with the Coinami project. Below, we provide a list of open problems and future directions.

\subsection{System-wide difficulty level} 
Currently, the Coinami network does not dynamically adjust the difficulty level in a way similar to Bitcoin. However, a reliable and self-regulating cryptocurrency system must limit the newly minted blockchains to prevent inflation. It is fairly straightforward in the Bitcoin protocol, where basically the difficulty level is determined by an integer, and changes to this number make it harder (or easier) to calculate the nonce value. Since the Coinami network requires an external data source, the system-wide difficulty level adjustment is not straightforward.

The first solution that comes to mind could be simply increasing the size of the assignment therefore increasing the time required to complete the task. However, this would mean that the assignment may need to be composed of billions of reads from multiple full genomes, which will incur a heavy load on storage and transfer, and eventually cause starvation due to the time-out mechanism that is designed to prevent deadlocks. A more sophisticated solution may be adding counters to the token and adjust the difficulty level ($D$). The authority may simply increment the counter and sign/finalize the blockchain only when 
the miner finishes $D$ assignments. This would also require the miner to send a pre-existing but not yet signed token to the authority, or if no such token is sent, the authority will create a new token with a counter initialized to 1. Alternatively, the authority may keep a database of miners and their token counters. Another possibility is that the authority sends out $D$ assignments to the miners, but sign only one of the randomly selected solutions. This would ensure $D$ assignments are completed, but only one block is validated from solutions. However, this scheme needs to be carefully tuned to provide fairness among miners.

\subsection{Authority-specific difficulty} We are also planning to add authority-specific difficulty levels to Coinami.
The system will increase the difficulty for assignments from those authorities that signs substantially more (amount to be determined) tokens than other authorities.
This is to prevent an authority to take over the system, and also to provide fairness among authorities. In addition, due to the multi-centric nature of Coinami, a ``strong'' (i.e. frequent signer)
authority may essentially become a monopoly (compared to other authorities) 
and affect the value and integrity of the cryptocurrency\footnote{Recent history showed that economies should avoid creating entities that are too big to fail.}.

\subsection{Data transfer} 
The storage and therefore data transfer requirements for HTS data are very high. For example, the raw data (i.e. FASTQ files) of a genome sequenced at 40X coverage consumes approximately 150 GB of space even when compressed with {\tt gzip}. The file size for the result (i.e. BAM file) will be similar. The assignments in Coinami will not contain full genomes, however, the amount of data to be transferred between miners and authorities will still be large, which will impose difficulties for both the authority and the miner. As a future work, we plan to incorporate data-specific compression to ameliorate this problem. For example, the assignments will be compressed using a FASTQ compressor such as SCALCE~\cite{Hach2012}, and the resulting BAM file will be compressed using DeeZ~\cite{Hach2014a} or CRAMtools~\cite{Fritz2011}. Although they will add to the computational burden for the miner, they will also reduce the file sizes by $\geq$50\%, which will also help with the data transmission problem. We are also planning to add
 native SCALCE and DeeZ capabilities to the read mappers used in Coinami and to the authority processing. 
 To further improve data transfer speed, we plan to use a UDP-based data transfer protocol, such as UDT~\cite{Gu2007}. 
With a view to privacy, all communications between the miner and the authority will be encrypted.


\subsection{Generalized application interface} 
As the Coinami protocol separates the proof-of-work from the rest of the system (i.e. transactions), it is possible to modify it to solve other scientific problems that require substantial computational capabilities. However, its current implementation only supports HTS read mapping\footnote{In fact, specifically Illumina data.}.

We plan to make Coinami easier to port for solving different problems and to provide a plug-in mechanism for other researchers to use the Coinami network for their computational needs. For this purpose, we plan to develop interfaces that use Docker containers as the proof-of-work component. This will also make it possible to switch from a single-application multi-authority system to multi-application multi-authority system. We call these new type of authorities ``employers'', the miners ``employees'', and the root authority ``the ministry of labor''. The employees will be able to select the scientific problem to contribute to the solution of, and the tokens will be signed by the employer that provided the job. Note that the current BOINC platform works similarly, except for the cryptocurrency system. As a side note, this approach also may lead to interesting game theory problems\footnote{One wonders if there will be any strikes and layoffs.}.

\section{Discussion}
The amount of HTS data generated world-wide is ever-increasing, and it is expected to surpass other major domains of ``big data'': YouTube, Twitter, and Astronomy~\cite{Stephens2015}. 
Hundreds of thousands of genomes and exomes are sequenced every year, 
which creates a huge computational burden. Additionally, even when we focus only on human data, the reference genome receives an update every few years, causing additional computation cost as the pre-existing data need to be remapped to the new version of the reference. As sequencing in the clinical environments gain traction~\cite{Biesecker2009}, the volume of such data will only skyrocket,
not to mention shifting priorities and urgency imposed by clinical cases, leading to  ``analysis starvation'' for data with less priority. There are also efforts in place
to move from a linear reference genome to graph-based structures to build ``pan-genomes'' that represents whole populations~\cite{Nguyen2015}, and ``variant graphs'' to represent all previously known genomic variants (\url{https://github.com/ekg/vg}), which will likely prompt re-analysis of most existing data.

Currently HTS data analyses are performed using large scale clusters. 
Building and maintaining clusters are not easy tasks, and for most users certainly not feasible. Against this background the research community is witnessing a shift~\cite{Stein2015} 
to using dedicated academic clouds for genomics, such as the Embassy Cloud~\cite{Cook2016} at the European
Bioinformatics Institute (EBI), and multi-purpose commercial cloud platforms such as Amazon AWS. 
However, no clusters, cloud platforms, or data centers offer ``infinite'' capabilities, and with HTS data growing faster than the fruits of Moore’s Law, the need for alternative approaches to distribute some of the workload to as many computers as we can is obvious. Volunteer grid computing may help analyze low-priority data (i.e. remapping to new reference), however, it is also less likely for volunteers to dedicate necessary resources as HTS mapping is compute intensive, requires large storage space and network bandwidth to download raw data and upload results. It might be possible to merge cryptocurrencies with volunteer grid computing as an approach to help increase volunteer motivation.

Coinami provides a protocol and prototype implementation of such a combination of Bitcoin and BOINC. We show that it is possible to distribute read mapping work load to untrusted parties (i.e. volunteers or miners), while ensuring data privacy and preventing malicious users that may try to submit ``garbage'' results. However, the miners have to trust the root authority, which in turn must trust the middle level authorities.

It is unknown whether Coinami will be used in practice in the future. There are three  questions left unanswered yet. First: who shall act as the root authority? Probably, this is the easiest to answer, as a universally trusted research entity - such as the National Center for Biotechnology Information (NCBI) or EBI - may assume this role. The root authority only approves or rejects middle level authorities, therefore running
the root authority server does not induce computation overhead. The second question is more complicated: will any sequencing center join the system as a middle level authority? The authorities have to 
invest on computational capacity to generate assignments and to validate the results returned by miners. They also have to actually be willing to share the needed data. The data sharing problem can be easily solved if only publicly available pre-existing data is injected to the system for realignment purposes\footnote{There is, however, a risk that editors of a certain journal may call miners as \#researchparasites.}.
However,  the computational burden of running an authority server must be assessed to see if it is more feasible than performing the analyses in-house. Finally: will Coinami have the desired effect for volunteer motivation? This is probably the hardest one to answer, which will also depend on the ``popularity'' and value of Coinami-generated cryptocurrency. One possible approach to solve the ``economics'' problem may be fixed-rate conversion of Coinami blockchains with traditional currency, paid by the authority that issued the corresponding token. Another possibility is for the authorities 
to offer free service such as exome sequencing in 
exchange of Coinami blockchains. Or, Coinami blockchains may just remain as  ``bragging rights'' 
for {\it real} volunteers and science and technology enthusiasts.
Although less likely, people may deem Coinami blockchains to be valuable in-and-of themselves, and the Coinami project may even spearhead future developments to accelerate read mapping using field programmable gate arrays (FPGA) and application specific integrated circuits (ASIC).  

\section{Acknowledgments}
We thank O. Mutlu, V. Boeva, F. de la Vega, R. Durbin, R. Chikhi, C. Orhan and N. Lack for discussions and future work ideas about this project, and C.T. Ozturk for proofreading the manuscript.
We also thank M. Inan, E.B. Gulcan, and E. Akay for helping with the prototype implementation. This work was supported in part by an European Molecular Biology Organization 
Installation Grant (IG-2521) to CA.

\section{Author contributions}
AMI developed the initial concept. AMI and CA defined protocols. HIO, AG, AKS, and MYO. modified parts of the protocols and implemented the prototype. AMI, HIO, and CA wrote the manuscript.

    \bibliographystyle{abbrv}
    \bibliography{calkan}
\end{document}